\documentclass[pdflatex,sn-nature]{sn-jnl}

\usepackage{graphicx}%
\usepackage{amsmath,amssymb,amsfonts}%
\usepackage{xcolor}%
\usepackage{textcomp}%
\usepackage{booktabs}%

\begin{document}

\title{Deterministic and dynamic synthesis of optical skyrmions through an ultrathin multimode fibre}

\author[1]{\fnm{R. E.} \sur{Bautista Gonzalez}}

\author[1]{\fnm{R.} \sur{Mouthaan}}

\author[2]{\fnm{A.} \sur{McIntosh}}

\author*[3,1]{\fnm{K.} \sur{Dholakia}}\email{kd1@st-andrews.ac.uk}

\affil[1]{\orgdiv{Centre of Light for Life and School of Biological Sciences}, \orgname{Adelaide University}, \orgaddress{\state{SA} \postcode{5005}, \country{Australia}}}

\affil[2]{\orgdiv{School of Physics}, \orgname{UNSW Sydney}, \orgaddress{\city{Kensington}, \state{NSW} \postcode{2052}, \country{Australia}}}

\affil[3]{\orgdiv{SUPA, School of Physics and Astronomy}, \orgname{University of St Andrews}, \orgaddress{\street{North Haugh}, \city{St Andrews}, \postcode{KY16 9SS}, \country{UK}}}

\abstract{
%Topological light fields offer powerful opportunities for information encoding, transport, and manipulation, yet their deterministic generation in complex media remains an outstanding challenge. In highly multimode fibres, strong mode coupling and polarisation mixing obscure the precise control required to synthesise structured vectorial states, including optical skyrmions. Here, we demonstrate the deterministic generation and dynamic reconfiguration of optical skyrmions through an ultrathin multimode fibre. By combining vectorial transmission-matrix control with a phase-retrieval optimisation framework, we exploit the polarisation-resolved modal space to reconstruct user-defined vector fields with high fidelity at the distal facet. This approach enables the synthesis of Néel, Bloch, anti- and higher-order skyrmions, and bimerons, whose topological properties are confirmed through Stokes-vector measurements and the skyrmion number. Wavefront reprogramming allows transitions between distinct skyrmionic states. Our results establish multimode fibres as programmable platforms for topological light, with potential applications in minimally invasive imaging and quantum and optical communications.

Topological light fields offer powerful opportunities for information encoding, transport, and manipulation, yet their deterministic generation in complex scattering media remains an outstanding challenge. In highly multimode fibres, strong mode coupling and polarisation mixing can obscure the precise control required to synthesise structured vectorial states, including optical skyrmions. Here, we demonstrate the deterministic generation and dynamic reconfiguration of arbitrary optical skyrmions through an ultrathin multimode fibre. By combining vectorial transmission-matrix control with a phase-retrieval optimisation framework, we exploit the full polarisation-resolved modal space of the fibre to reconstruct user-defined vector fields with high fidelity at the distal facet. This approach enables the synthesis of Néel, Bloch, anti- and higher-order skyrmions, and bimerons. Their topological properties are confirmed through polarisation-resolved Stokes-vector measurements and quantitative determination of the skyrmion number. Wavefront reprogramming further provides dynamic control over polarisation textures, enabling reversible transitions between distinct skyrmionic states. Our results establish multimode fibres as programmable platforms for the generation of topological light and introduce a scalable route towards minimally invasive topological photonics, with potential applications in imaging, sensing, optical communications, and information processing in complex environments.
}

\maketitle

\section{Introduction}\label{sec:intro}

Skyrmions are topologically protected field configurations originally introduced in particle physics as stable, localised solutions of nonlinear field theories \cite{skyrme1962unified}. Since their introduction, skyrmionic textures have been observed across a wide range of physical systems, including magnetic materials \cite{muhlbauer2009skyrmion}, acoustic waves \cite{ge2021observation}, and structured optical fields \cite{shen2022generation}. In optics, skyrmions arise as vector light fields whose spatially varying polarisation maps continuously onto the Poincaré sphere, producing non-trivial topological textures characterised by an integer-valued skyrmion number \cite{shen2024optical,yang2025optical,peters2026extracting}. Their unique combination of topological protection and high-dimensional vectorial structure has stimulated growing interest in applications ranging from information encoding and optical communications to sensing, particle manipulation, and structured-light engineering \cite{shen2024optical,yang2025optical,shen2025topological,wang2024topological}.

Recent advances have enabled the generation of increasingly sophisticated optical skyrmions using free-space beam shaping \cite{shen2021supertoroidal,shen2022generation,shen2024nondiffracting}, metasurfaces and nanophotonic devices \cite{he2024optical}, laser cavities \cite{kerridge2024optical}, and plasmonic systems \cite{shen2024plasmonic,shen2024optical}. Among these approaches, spatial light modulators (SLMs) provide exceptional flexibility by enabling programmable control over the amplitude, phase, and polarisation \cite{shen2022generation}. However, most demonstrations remain confined to controlled free-space environments. Extending deterministic skyrmion generation to complex propagation media is essential for translating topological light into practical technologies operating in realistic and inaccessible environments.

Multimode optical fibres (MMFs) provide a particularly compelling platform for this transition. Their large modal capacity offers access to a high-dimensional optical state space, while their ultrathin cross-sections enable minimally invasive light delivery deep within confined or scattering environments \cite{vcivzmar2012exploiting,stibuurek2023110,turtaev2018high,vasquez2018subcellular}. Such capabilities are attractive not only for endoscopic imaging and sensing but also for topological and quantum information technologies. The large number of supported modes naturally enables high-capacity communications through mode-division multiplexing \cite{shi2026tailoring,shi2025generation}, while the skyrmion number provides a robust topological degree of freedom that can encode information resilient to perturbations of the underlying polarisation texture \cite{peters2026extracting,wang2024topological,guo2026topological}. Furthermore, recent demonstrations of non-local optical skyrmions generated using entangled photons suggest opportunities for topology-enhanced quantum communication and sensing \cite{ornelas2024nonlocal,ornelas2025topological}. Realising these opportunities, however, requires deterministic control over both the spatial and polarisation degrees of freedom of light at the fibre output.

Achieving such control remains challenging because light propagation through a highly MMF is governed by strong modal and polarisation mixing. These effects scramble the optical field, obscuring any direct relationship between the incident and transmitted wavefronts. Transmission-matrix approaches have transformed this challenge into an opportunity by providing a complete linear description of light transport through complex fibres, enabling deterministic focusing, beam shaping, holography, and image projection \cite{vcivzmar2011shaping,popoff2010measuring,mouthaan2025generating,ploschner2015seeing,cao2023controlling}. More recently, vectorial transmission-matrix (TM) methods have revealed that polarisation mixing can also be measured and controlled, enabling the reconstruction of arbitrary polarisation states at selected output locations and across extended optical fields \cite{mounaix2019control,xiong2018complete,zhang2026high,zhuang2024dynamic,gomes2022near}. Despite these advances, the deterministic generation of topological vector fields, and optical skyrmions in particular, through highly MMF has remained elusive.

\begin{figure}[!ht]
\centering
\includegraphics[width=1\linewidth]{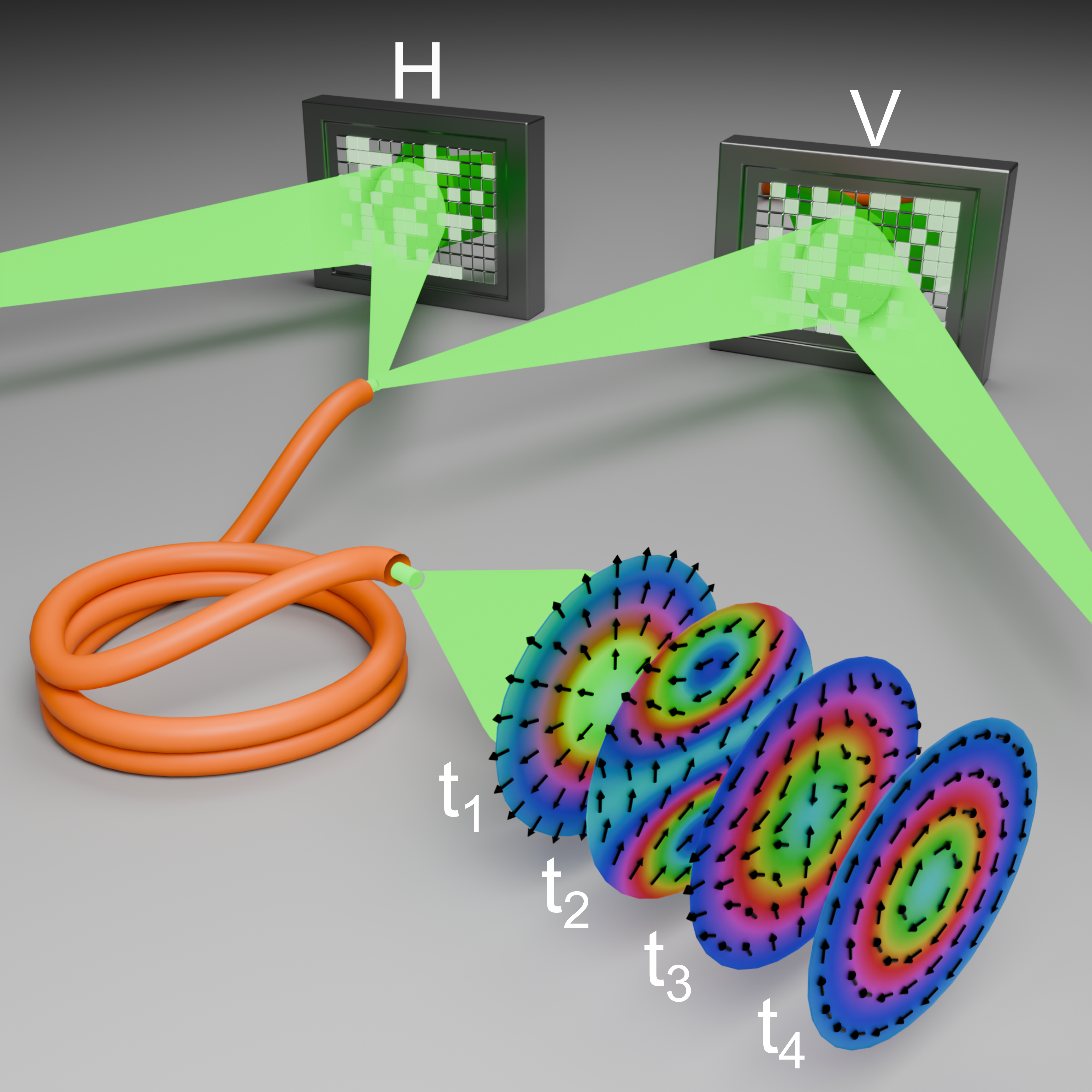}
\caption{\textbf{Dual-polarisation wavefront shaping.} Two SLMs, controlling two independent polarisations, are used to modulate the input field of the MMF. This can be used to achieve the deterministic and dynamic generation of skyrmions once the TMs for both polarisations are characterised.}
\label{fig:concept}
\end{figure}

Here, we demonstrate that the full vectorial complexity of light propagation in a highly MMF can be harnessed to deterministically generate and dynamically control optical skyrmions at the distal fibre facet (Fig. \ref{fig:concept}). By combining vectorial transmission-matrix measurements with a phase-retrieval optimisation framework, we achieve programmable synthesis of a diverse family of skyrmionic states, including Néel, Bloch, anti-skyrmion, bimeron, and higher-order textures, following propagation through a long and deformed MMF. Polarisation-resolved measurements enable complete reconstruction of the Stokes-vector field and quantitative evaluation of the skyrmion number, confirming high-fidelity realisation of the targeted topologies. Previous fibre-based approaches have relied on metasurfaces integrated at the tip of a single-mode fibre \cite{he2024optical}, which require complex fabrication and provide only static functionality, or have relied on few-mode fibres \cite{shi2025generation,shi2026tailoring}, where the limited modal degrees of freedom constrain the range and flexibility of accessible skyrmionic states. Specialty ring-core fibres guide topologically confined orbital angular-momentum modes that resist intermodal coupling \cite{ma2023scaling, peters2025guided, ashok2026robust, peters2026combining}. In these guided approaches the skyrmion number is preserved, but the texture itself can drift with the relative phase between the constituent modes \cite{shi2026tailoring, peters2026combining}. In contrast, our approach enables the deterministic generation and dynamic switching of a broad library of topological textures using a standard commercial MMF, without the need for specialised fibre or nanophotonic fabrication. Beyond representing a major advance in optical skyrmion generation, our results establish MMFs as programmable platforms for topological light synthesis, transforming a medium traditionally associated with modal scrambling into a versatile tool for vectorial field engineering. This capability opens new opportunities in topological photonics, minimally invasive imaging and sensing, and future classical and quantum information technologies.

\section{Results}\label{sec:results}

\begin{figure}[!ht]
\centering
\includegraphics[width=1\linewidth]{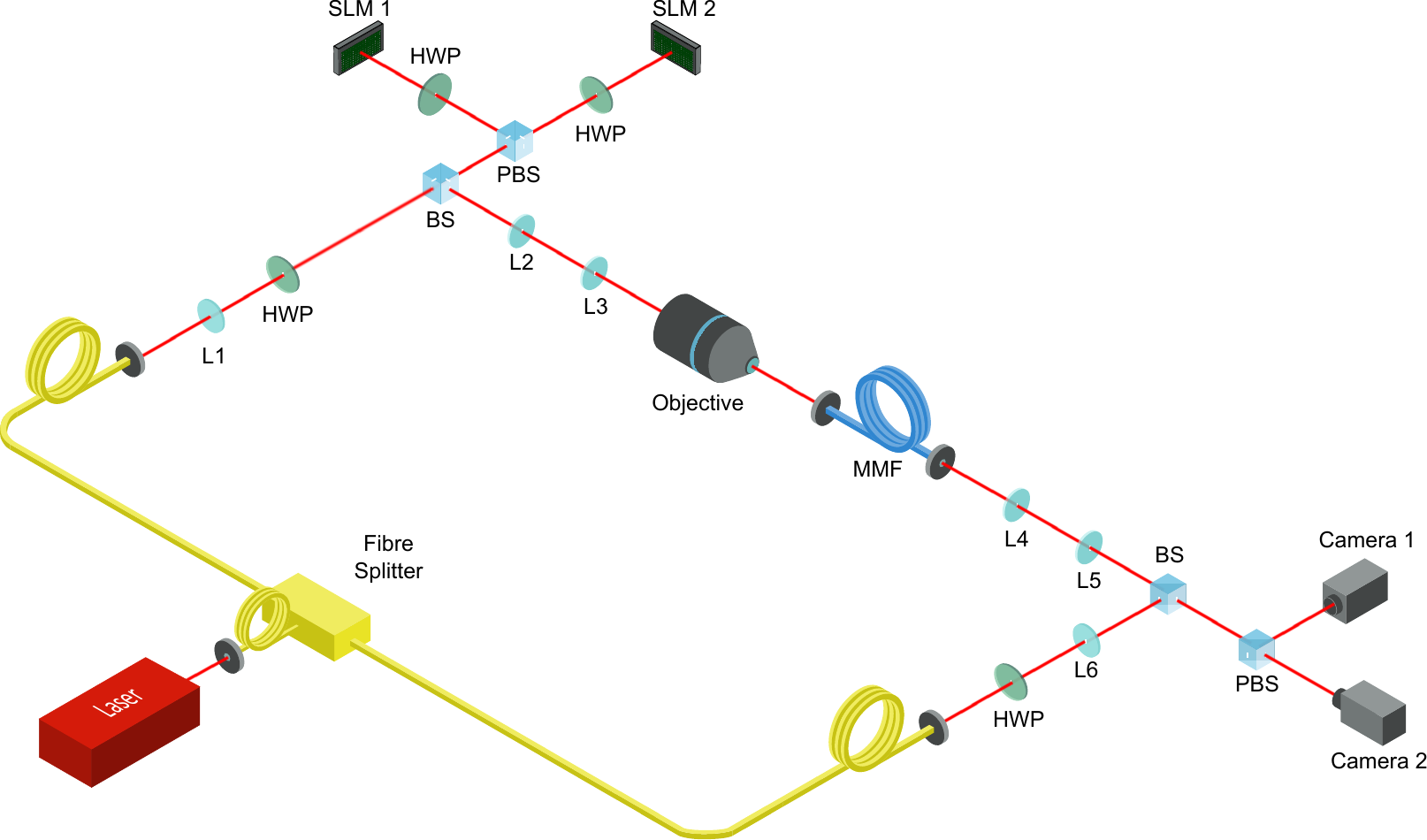}
\caption{\textbf{Experimental setup for dual-polarisation wavefront shaping through a MMF.} L denotes lenses, HWP denotes half-wave plates, PBS denotes polarising beam splitter, and BS denotes beam splitter. The components used are detailed in Supplementary Table~1.}
\label{fig:expsetup}
\end{figure}

The experimental setup and procedure for characterising the vectorial transmission matrix of a 1~m long OM1 multimode graded-index fibre is described in detail in the Methods and illustrated in Fig. \ref{fig:expsetup}. Two spatial light modulators are employed to control the orthogonally polarised MMF input fields, and off-axis holography is used to characterise the orthogonally polarised MMF output fields. Separate transmission matrices are characterised for each combination of input and output polarisations. A differential algorithm was used to correct for phase drift and unwanted scattered light over the course of these measurements \cite{mcintosh2026differentially}, ensuring that a highly accurate estimate of the transmission matrix is obtained. A measurement of the phase offset between the various polarisation combinations is taken \cite{mouthaan2022robust}, allowing the vectorial transmission matrix to then be constructed. This vectorial transmission matrix relates 2048 input modes, corresponding to macropixels across both SLMs, to 2048 output modes, corresponding to a meshgrid of 2048 pixels distributed across both output polarisations. 

Once the vectorial transmission matrix has been characterised, optimised holograms for projecting a given optical skyrmion are synthesised using a bespoke phase-retrieval algorithm. In contrast to direct transmission-matrix inversion approaches, phase retrieval algorithms iteratively refine the input field by alternating between the input and output domains while enforcing the physical constraints of the system \cite{gerchberg1972practical,fienup1982phase,yang1994gerchberg,mouthaan2025generating}. At the input plane, the phase-only modulation constraint imposed by the SLMs is enforced. At the output plane, the calculated field is constrained to match the desired target field. The optimisation is iterated until convergence of the output-field fidelity. Here, a bespoke weighted Yang--Gu (YG) phase retrieval algorithm \cite{mouthaan2025generating, yang1994gerchberg} is employed, which allows amplitude, phase and polarisation to be constrained at the output and is optimised for through-fibre projections by accounting for the non-unitary nature of the transmission matrix. The resulting holograms are displayed on the two SLMs, enabling deterministic generation of the desired vector field at the distal facet of the MMF.

\begin{figure}[!ht]
\centering
\includegraphics[width=0.7\linewidth]{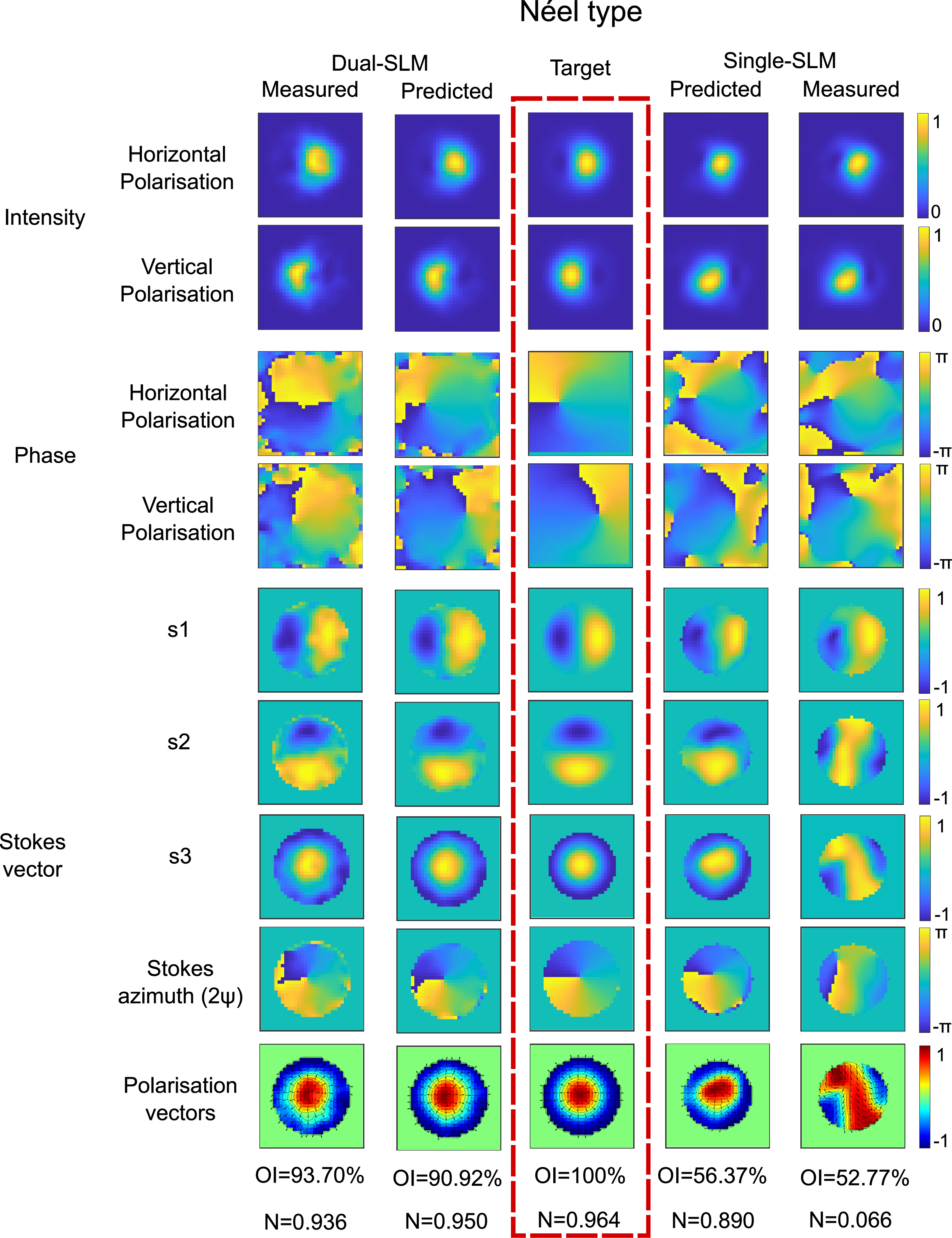}
\caption{\textbf{Comparison of single-polarisation and dual-polarisation wavefront shaping for the generation of a N\'eel type optical skyrmion through an MMF.} The central column, highlighted by the red dashed rectangle, shows the target field, while the adjacent columns show the fields predicted using the experimentally measured TMs and the outer columns show the corresponding experimental measurements. Results are presented for the intensity and phase of the horizontal and vertical polarisation components, followed by the reconstructed normalised Stokes parameters $s_1$, $s_2$, and $s_3$, the Stokes azimuth 2$\Psi$ = atan2(s2, s1), and the spatial polarisation-vector texture, where $\Psi$ is the orientation of the polarisation ellipse. For dual-polarisation control, the predicted and measured fields closely reproduce the target N\'eel-type texture, the overlap integral (OI) with respect to the target field is $90.92\%$ for the predicted field and $93.70\%$ for the measured field with respect to the predicted field. The corresponding skyrmion numbers are $N_{\mathrm{sk}}=0.950$ and $0.936$, compared with the target value of $N_{\mathrm{sk}}=0.964$. In contrast, single-polarisation control produces a less faithful reconstruction, with respect to the target field is $56.37\%$ for the predicted field and $52.77\%$ for the measured field with respect to the predicted field. Although the predicted single-polarisation field retains a skyrmion number of $N_{\mathrm{sk}}=0.890$, the experimentally measured field yields $N_{\mathrm{sk}}=0.066$, indicating a substantial loss of the designed topological structure. These results demonstrate the importance of simultaneous control of both polarisation inputs for the robust experimental synthesis of optical skyrmions through the MMF.}
\label{fig:singleSLMvstwoSLM}
\end{figure}

We commence by generating a Néel-type optical skyrmion, requiring simultaneous control of the spatially varying amplitude, phase, and polarisation at the fibre output. The mathematical definition of the Néel-type optical skyrmion, as well as other skyrmions, is introduced in the Methods. This formulation allows the target field to be defined on the output meshgrid, illustrated in the central, highlighted column of Fig. \ref{fig:singleSLMvstwoSLM}. Necessary discretisation on the meshgrid causes a reduction in the skyrmion number from an ideal 1 to $0.964$, which effect could be mitigated by using an MMF supporting a larger number of modes. The weighted YG algorithm is used to obtain holograms for display on the two SLMs. The vectorial transmission matrix can be used to simulate projection through the fibre, yielding the predicted fields. Displaying the holograms on the SLMs, meanwhile, provides the experimentally measured vector fields. Both are shown on the left-hand side of Fig. \ref{fig:singleSLMvstwoSLM}. The similarity between the target and predicted fields, based on the overlap integral (OI), was $90.92\%$, while the similarity between the predicted and experimentally measured fields was $93.70\%$. An experimental skyrmion number of $0.936$ was obtained, approaching the skyrmion number for the target field of $0.964$.

We compare this initial dual-polarisation result to the case where only a single input polarisation is controlled. A new transmission matrix is calculated, corresponding to a single input polarisation but capturing both output polarisations. Additionally, the number of input modes is increased to 4096. As before, the weighted YG algorithm is used to generate a hologram, with the projection results now shown on the right-hand side of Fig. \ref{fig:singleSLMvstwoSLM}. In this case, the similarity between the target and predicted fields was $56.37\%$, the similarity between the predicted and experimentally measured fields was $52.77\%$, and an experimental skyrmion number of $0.066$ was obtained. Despite employing twice the number of input degrees of freedom, the single polarisation approach yielded poor results, and a skyrmion is not obtained. Practically, it is observed that while the intensity profile of skyrmion is correctly projected, the target phase profile is not faithfully recreated. The superior performance of the dual-polarisation system demonstrates that independent control of the both orthogonal input polarisation channels is considerably more important than simply increasing the number of controlled spatial modes.

\begin{figure}[p]
\centering
\includegraphics[width=0.6\linewidth]{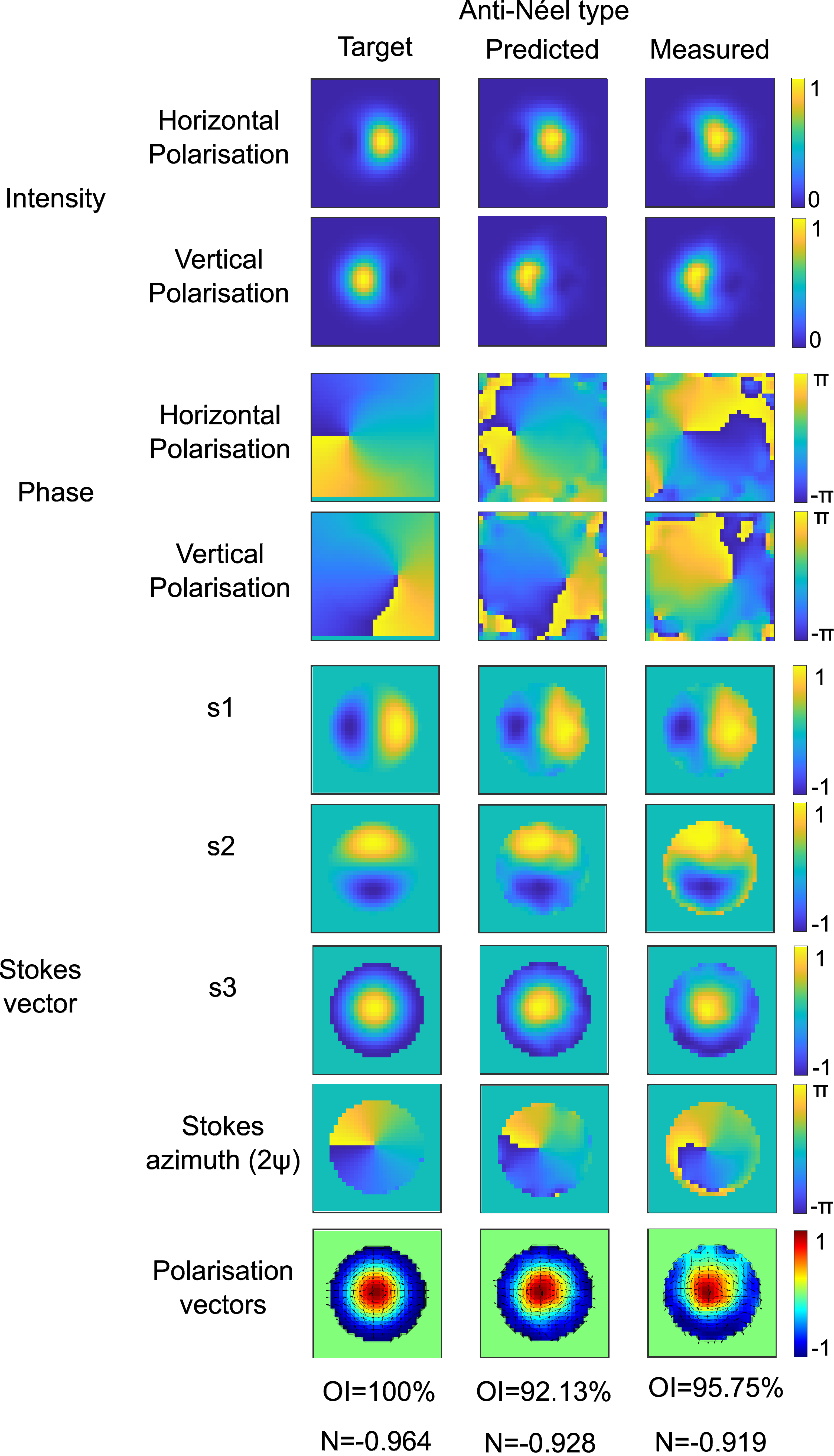}
\caption{\textbf{Generation and experimental characterisation of an anti-N\'eel type optical skyrmion through dual-polarization wavefront shaping.} The columns show the target field, the field predicted from the measured TM, and the experimentally measured field at the fibre output. The first four rows show the intensity and phase distributions of the horizontal and vertical polarization components. The reconstructed complex fields are used to calculate the normalised Stokes parameters $s_1$, $s_2$, and $s_3$, together with the corresponding Stokes azimuth 2$\Psi$ and spatial polarisation-vector texture. The characteristic anti-N\'eel polarisation texture is reproduced in both the predicted and measured fields. The overlap integral (OI) is $92.13\%$ for the predicted field with respect to the target field and $95.75\%$ for the measured field with respect to the predicted field. The corresponding skyrmion numbers are $N_{\mathrm{sk}}=-0.964$, $-0.928$, and $-0.919$ for the target, predicted, and measured fields, respectively, demonstrating preservation of the designed anti-N\'eel type topological structure.}
\label{fig:AntiNeelskyrmion}
\end{figure}

\begin{figure}[!ht]
\centering
\includegraphics[width=0.6\linewidth]{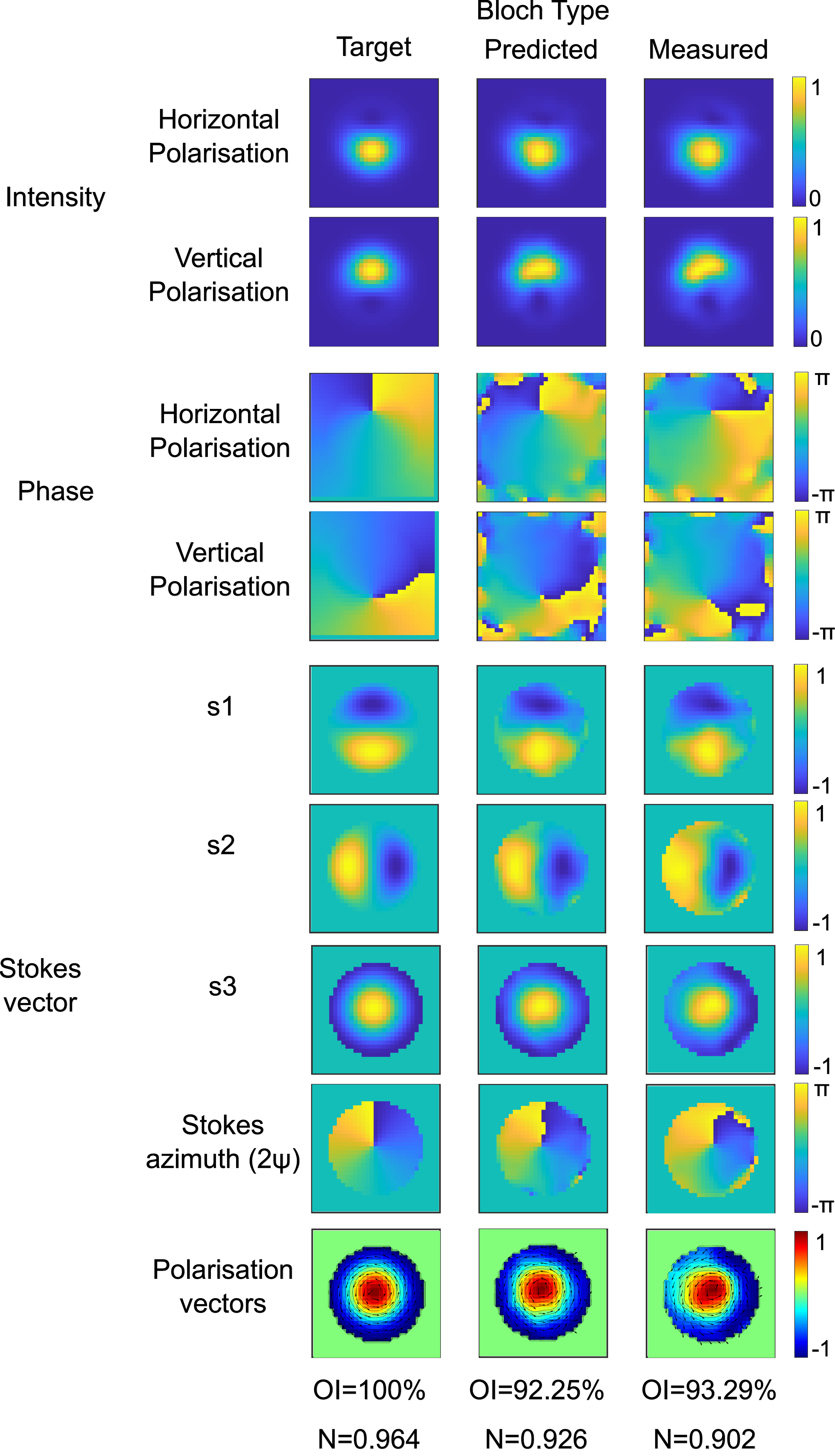}
\caption{\textbf{Generation and experimental characterization of a Bloch type optical skyrmion through dual-polarisation wavefront shaping.} The columns show the target field, the field predicted from the measured TM, and the experimentally measured field at the fibre output. The first four rows show the intensity and phase distributions of the horizontal and vertical polarisation components. The reconstructed complex fields are used to calculate the normalised Stokes parameters $s_1$, $s_2$, and $s_3$, together with the corresponding Stokes azimuth 2$\Psi$ and spatial polarisation-vector texture. In contrast to the radial texture of the N\'eel-type skyrmion, the Bloch-type state exhibits the characteristic azimuthal rotation of the polarisation vectors around the skyrmion core. The overlap integral (OI) with respect to the target field is $92.25\%$ for the predicted field and $93.29\%$ for the measured field with respect to the predicted field. The corresponding skyrmion numbers are $N_{\mathrm{sk}}=0.964$, $0.926$, and $0.902$ for the target, predicted, and measured fields, respectively, confirming preservation of the designed Bloch-type topology.}
\label{fig:Blochskyrmion}
\end{figure}

\begin{figure}[!ht]
\centering
\includegraphics[width=1\linewidth]{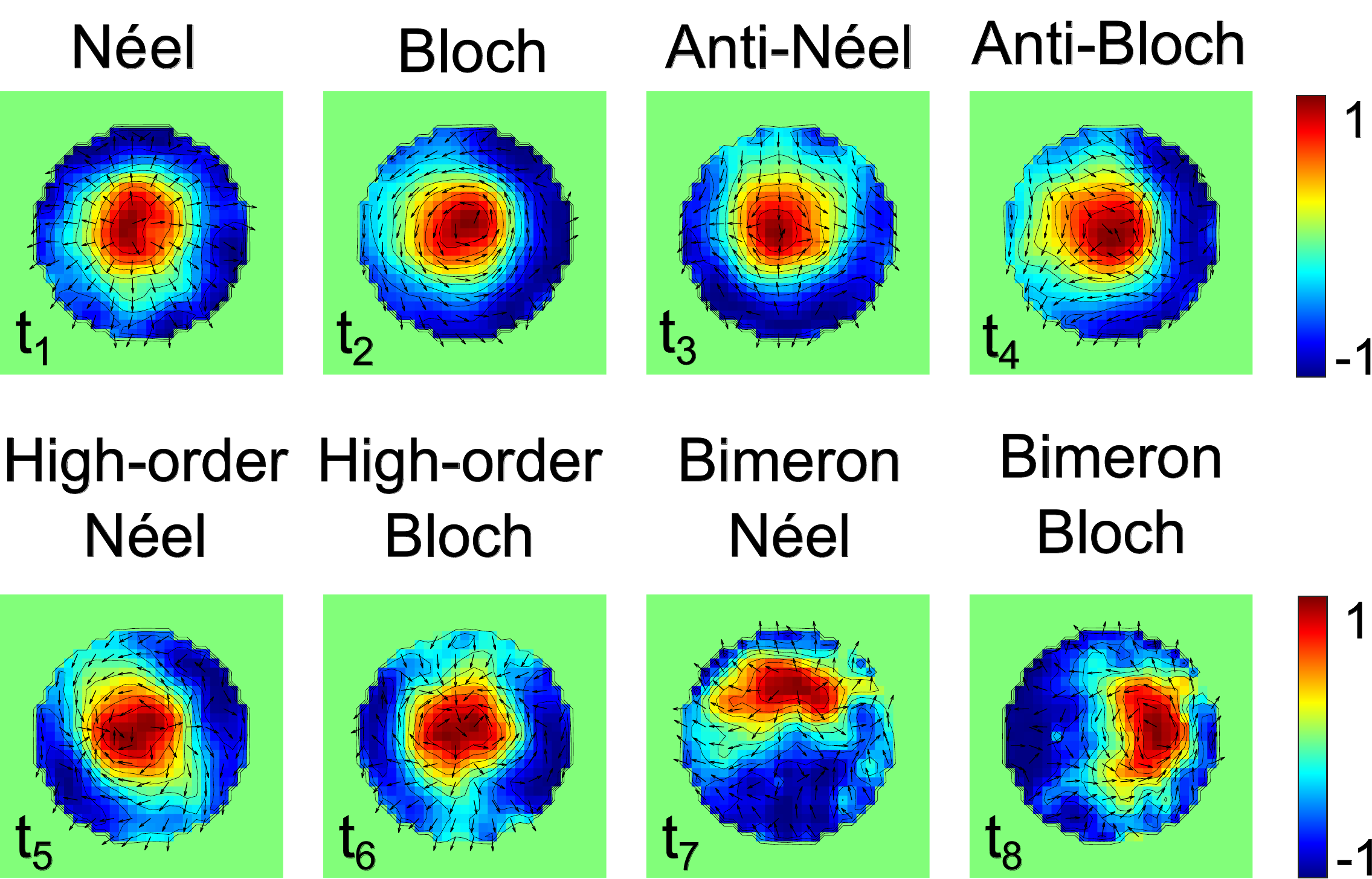}
\caption{\textbf{Consecutive generation of optical skyrmions through dual-polarisation wavefront shaping at the distal facet of an ultrathin MMF.} Eight distinct skyrmion types were projected sequentially at the distal facet of the MMF at times $t_1$--$t_8$ by updating only the phase masks displayed on the two SLMs, using a single vectorial transmission-matrix calibration and without any realignment of the optical system: N\'eel ($t_1$), Bloch ($t_2$), anti-N\'eel ($t_3$), anti-Bloch ($t_4$), higher-order ($m=2$) N\'eel ($t_5$) and Bloch ($t_6$), and N\'eel ($t_7$) and Bloch type ($t_8$) bimerons. Colour shows the experimentally measured normalised Stokes parameter ($s_3$), and black arrows indicate the orientation of the in-plane Stokes components ($s_1$, $s_2$). The skyrmion textures map the beam centre and periphery onto opposite poles of the Poincar\'e sphere, with the in-plane polarisation winding radially (N\'eel) or azimuthally (Bloch), in the opposite sense for the anti-skyrmions, and twice around the core for the higher-order states. The bimeron textures instead comprise two merons of opposite polarity. The full sequence was acquired in less than 120 seconds (this includes the calculation of the holograms), demonstrating dynamic, software-defined switching between topologically distinct states through the same fibre.}
\label{fig:skyrmionstime}
\end{figure}

Having established deterministic vectorial control, we next demonstrate the generation of a diverse range of topological polarisation textures. Figures \ref{fig:AntiNeelskyrmion} and \ref{fig:Blochskyrmion} present experimentally generated anti-Néel type and Bloch type optical skyrmions at the fibre output, respectively. Additional projections including anti-Bloch type, higher-order Néel type ($m=2$), higher-order Bloch type ($m=2$), bimeron-Néel type, and bimeron-Bloch type configurations are shown in Supplementary Figs.~1--5. These projections are performed by characterising the transmission matrix once and then projecting the various skyrmions in sequence. Including the time to calculate the holograms using the weighted YG algorithm, it takes less than 120 seconds to project all eight optical skyrmions illustrated in Fig. \ref{fig:skyrmionstime}. Should the holograms be pre-calculated, it will become possible to dynamically change the projected skyrmion with a refresh rate determined by the SLM (typically 30~Hz). For all eight skyrmion types, the similarity between the target and predicted fields exceeds $90.78\%$, while the similarity between the predicted and experimentally reconstructed fields also remains above $91.57\%$. The reconstructed amplitude, phase, and polarisation distributions closely match the target vector fields, demonstrating that the measured vectorial transmission matrix combined with the weighted YG algorithm enables deterministic synthesis of complex topological vector fields through a highly multimode optical fibre. The reconstructed Stokes vector fields were used to calculate the skyrmion number of the target, predicted, and experimentally reconstructed optical fields. The experimentally measured skyrmion numbers closely match the values for the target fields, with relative errors below $8.1\%$ (mean $4.95\%$). These results demonstrate deterministic generation of optical skyrmions with high topological fidelity at the distal facet of the MMF. By moving to highly multimode MMFs, the range of skyrmions that can be dynamically projected using a single system has been greatly extended. The range of skyrmions that can be projected through a given fibre is ultimately restricted by the number of modes available, but skyrmion types of higher orders beyond those considered here can be projected if a fibre with a larger mode capacity were to be used.

\section{Discussion}\label{sec:discussion}

The deterministic generation of optical skyrmions through a highly multimode optical fibre has been demonstrated using dual-polarisation wavefront shaping. By combining vectorial transmission matrix characterisation with the weighted YG phase-retrieval algorithm, we achieved independent control of two orthogonal polarisation channels at the distal fibre facet, enabling the synthesis and dynamic reconfiguration of eight types of optical skyrmions, including Néel, Bloch, anti-skyrmion, bimeron, and higher-order textures. The close agreement between designed, predicted, and experimentally reconstructed fields, together with quantitative recovery of the skyrmion number, confirms that complex topological polarization structures can be faithfully generated following propagation through a strongly mode-mixing optical system.

These results substantially extend previous efforts to generate optical skyrmions through optical fibres. Earlier approaches  relied on metasurface-functionalised fibre probes \cite{he2024optical}, which require specialised nanofabrication and provide only static functionality, or on fibres in which mode mixing is suppressed by design: few-mode fibres \cite{shi2025generation,shi2026tailoring}, where the limited modal dimensionality constrains the diversity of accessible topological states, or specialty ring-core fibres supporting topologically confined modes \cite{ma2023scaling, peters2025guided, ashok2026robust, peters2026combining}. The latter achieve long propagation distances, high skyrmion numbers and intrinsic resilience to intermodal coupling, phase noise and depolarisation. This makes them well suited to long-haul topological communication. However, because the skyrmion is carried by a fixed pair of fibre eigenmodes, its texture is set by the modal profiles and can drift with the intermodal phase, with only the skyrmion number preserved \cite{shi2026tailoring, peters2026combining}. In contrast, our approach employs a standard, ultrathin commercial MMF and employs software-defined wavefront control to achieve deterministic generation of prescribed textures, including their helicity and radial profile. Our approach enables rapid switching between a broad library of skyrmionic states. Furthermore, whereas the guided approaches above operate at telecommunication wavelengths (1310--1550~nm) \cite{shi2026tailoring, ashok2026robust, peters2026combining}, operation at a much broader range of wavelengths, notably visible wavelengths, as we demonstrate here, enables direct compatibility with a wide range of fluorescent probes. This opens a route to topological light delivery in biological and biomedical settings where infrared wavelength fibre based implementations are not compatible.

Several practical challenges remain. Transmission matrix acquisition currently requires approximately 24 minutes, limiting deployment in dynamically changing environments. However, substantial reductions in calibration time should be achievable through high-speed modulation technologies such as digital micromirror devices \cite{ayoub2021high, popoff2026practical} or fast phase-only spatial light modulators \cite{rocha2024fast}, as well as emerging compressive \cite{li2021compressively} and machine-learning-assisted \cite{fan2022deep, zhang2025fpga} calibration strategies. The approach also inherits the well-known sensitivity of MMFs to bending, twisting, and environmental perturbations. Nevertheless, recent advances in robust fibre design, real-time recalibration, and perturbation-aware wavefront shaping suggest viable strategies for overcoming these limitations and extending operation beyond controlled laboratory conditions \cite{cao2023controlling,boonzajerflaes2018robustness,turtaev2018high}. In addition, the requirement for distal access during calibration can potentially be removed through reflective characterisation schemes based on metasurface reflector stacks or related approaches \cite{gordon2019characterizing}.

The physical footprint of the platform further enhances its potential future impact. The MMF used here is approximately an order of magnitude thinner than conventional multicore fibre bundles used for endoscopic light delivery and is comparable in diameter to fibres already employed in minimally invasive endomicroscopy \cite{vcivzmar2012exploiting,stibuurek2023110,turtaev2018high,vasquez2018subcellular}. The combination of a hair-thin geometry with deterministic control over topological vector fields creates opportunities for delivering structured light into previously inaccessible environments. Such capabilities could benefit deep-tissue imaging, in vivo microscopy, neuroscience, photostimulation, and minimally invasive sensing, where precise control of optical fields must be achieved within highly constrained volumes.

More broadly, our results demonstrate that a strongly mode-mixing MMF can be transformed from a medium traditionally associated with modal scrambling into a programmable platform for topological light synthesis. Importantly, the approach requires no bespoke fibre fabrication and can be translated to new wavelength regimes through the straightforward selection and calibration of commercially available fibres. The demonstrated skyrmion families, distinguished by their topological charge, helicity, and texture, may therefore provide a reconfigurable and topologically robust alphabet for optical information encoding and communications. Looking further ahead, the framework naturally extends to quantum regimes. Because the vectorial transmission matrix fully describes the linear action of the fibre on the joint polarisation-spatial mode space, the same formalism applies at the single-photon level, opening opportunities for transmitting and manipulating non-local quantum skyrmions and other high-dimensional topological quantum states through ultrathin MMFs \cite{ornelas2024nonlocal,ornelas2025topological}.

In conclusion, this work establishes a new paradigm for topological photonics in complex optical systems. By enabling deterministic, reconfigurable generation of optical skyrmions through an ultrathin MMF, it bridges wavefront shaping, structured light, and topological photonics, while providing a versatile platform for future applications in deep-tissue imaging, sensing, communications, and quantum information science.

\section{Methods}\label{sec:methods}

\subsection*{Experimental setup}

The experimental system, shown in Fig. \ref{fig:expsetup}, is based on a highly multimode graded-index optical fibre (OM1, 62.5/125 $\mu m$ core/cladding diameter, 0.275 NA, FC/PC – FC/PC graded-index patch cable, 1 m long) supporting approximately 2576 guided spatial modes across two polarisation states at the operating wavelength of 532~nm. Light from a continuous-wave laser source (Coherent Sapphire, 532 nm) is divided into two paths using a fibre beam splitter. One path serves as the signal arm, while the other provides the reference beam for holographic measurements. In the signal arm, the beam is split into two orthogonal polarisation components using a polarising beam splitter (PBS), with each polarisation independently modulated by an SLM (Meadowlark E19X12-400-700 and Hamamatsu X10468-01). The two SLMs provide independent control over the orthogonal input polarisation states, enabling full vectorial modulation of the optical field at the fibre input. The modulated beams are then recombined and coupled into the MMF through an objective lens (20X Olympus UPlanFL N Objective, 20x/0.50, 2.1 mm WD Olympus, Tokyo, Japan). At the distal facet, the transmitted field is separated into its orthogonal polarisation components using a second PBS and simultaneously recorded by two identical cameras (FLIR Blackfly S BFS-U3-16S2M, Oregon, USA). The complex amplitude and phase of each polarisation component are reconstructed by off-axis digital holography using the reference beam from the second arm of the fibre beam splitter.

\subsection*{Vectorial transmission matrix measurement}

The MMF is treated as a linear optical system relating the complex input field generated by the SLMs $\mathbf{E}^{\mathrm{SLM}}$ to the output field measured by the camera $\mathbf{E}^{\mathrm{Cam}}$ by means of the transmission matrix $\mathbf{T}$. Subscripts are used to denote the input and output polarisation states. This allows this relationship to be expressed as

\begin{equation}
\begin{pmatrix}
\mathbf{E}^{\mathrm{Cam}}_1\\
\mathbf{E}^{\mathrm{Cam}}_2
\end{pmatrix}
=
\begin{pmatrix}
\mathbf{T}_{11} & \mathbf{T}_{21}\\
\mathbf{T}_{12} & \mathbf{T}_{22}
\end{pmatrix}
\begin{pmatrix}
\mathbf{E}^{\mathrm{SLM}}_1\\
\mathbf{E}^{\mathrm{SLM}}_2
\end{pmatrix}.
\label{eq:TM}
\end{equation}

The transmission matrix is characterised using a differential sampling method \cite{mcintosh2026differentially}, whereby each input mode is characterised using a differential pair of projections. This allows the effects of phase drift and common mode noise to be compensated for, giving a high-quality estimate of the final transmission matrix. The measurement is first performed for one SLM, allowing $\mathbf{T}_{11}$ and $\mathbf{T}_{12}$ to be obtained; and then done for the second SLM, allowing $\mathbf{T}_{21}$ and $\mathbf{T}_{22}$ to be obtained. A slow temporal phase drift is observed between the input fields originating from the two SLMs, which is characterised and accounted for using the approaches outlined in \cite{mouthaan2022robust}. This then allows the complete vectorial transmission matrix to be constructed.

The vectorial transmission matrix exhibits significantly poorer temporal stability compared to its single polarisation counterpart. This may, in part, be caused by relative instabilities between the two SLMs. Additionally, it is hypothesised that this instability also stems from the fact that small perturbations quickly lead to fibre modes being scattered into the corresponding degenerate mode of the opposite polarisation. Controlling a large number of modes requires a proportionally longer transmission matrix acquisition, during which mechanical and thermal drifts progressively reduce the validity of the measured transmission matrix. The impact of this phenomenon was reduced by employing only 1024 input modes for each input polarisation, allowing the fibre to be characterised in approximately 24 minutes. The vectorial fibre transmission matrix acquired in this manner was found to provide stable and reliable performance for one hour, in contrast to the single polarisation transmission matrix which was found to be stable for one day.

\subsection*{Formulation of optical skyrmions}
Optical skyrmions are defined by prescribing the spatial distribution of the polarisation state across the plane \cite{gao2020paraxial, shen2024optical, cisowski2023building}. The target polarisation texture is first defined through the normalised Stokes vector field, from which the corresponding Jones vector is obtained and used as the target field for wavefront shaping.
 
For a prescribed polarisation texture, the corresponding normalised Jones vector is obtained through a inverse Hopf map
 
\begin{equation}
\mathbf{E}(r,\phi)
=\begin{pmatrix}
\mathbf{E}_{R}(r,\phi)\\[1ex]
\mathbf{E}_{L}(r,\phi)
\end{pmatrix}
=
\begin{pmatrix}
\cos\!\left(\dfrac{f(r)}{2}\right)\\[1ex]
e^{-i(m\phi+\gamma)}
\sin\!\left(\dfrac{f(r)}{2}\right)
\end{pmatrix},
\label{eq:hopf}
\end{equation}
 
\noindent where $r$ and $\phi$ denote the radial and azimuthal coordinates, respectively, and $f(r)$ is the radial profile, $m$ is the topological order and $\gamma$ is the helicity \cite{shen2022generation, cisowski2023building}. The polarisation texture of an optical skyrmion is then described by the normalised Stokes vector
 
\begin{equation}
\mathbf{s}(r,\phi)
=
\begin{pmatrix}
s1(r,\phi)\\[1ex]
\;
s2(r,\phi)\\[1ex]
\;
s3(r,\phi)
\end{pmatrix}
=
\begin{pmatrix}
\sin f(r)\cos(m\phi+\gamma)\\[1ex]
\;
\sin f(r)\sin(m\phi+\gamma)\\[1ex]
\;
\cos f(r)
\end{pmatrix},
\label{eq:stokes_target}
\end{equation}
 
The radial profile $f(r)$ is chosen as a linear ramp,

\begin{equation}
f(r) =
\begin{cases}
\pi r/R, & r < R \\
\pi, & r \geq R,
\end{cases}
\label{eq:f_profile}
\end{equation}

\noindent analogous to the linear approach used to describe magnetic skyrmion profiles \cite{kiselev2011chiral}. It is a continuous, non-decreasing function satisfying $f(0)=0$ and $f(r)=\pi$ for $r \geq R$, where $R$ is the radius at which the polarisation reaches the opposite pole of the Poincar\'e sphere. In our experiments $R$ is chosen to lie within the controllable region of the fibre output. The polarisation therefore evolves continuously from one pole of the Poincar\'e sphere at the skyrmion centre to the opposite pole at the periphery of the beam ($r \geq R$), reaching this pole at a finite radius rather than asymptotically as $r \rightarrow \infty$, hence defining a continuous mapping from a bounded transverse aperture onto the Poincar\'e sphere. 
 
The parameter $m$ determines the topological order of the skyrmion, while the helicity $\gamma$ specifies the orientation of the in-plane polarisation. For a Néel-type skyrmion, $\gamma=0$ and the in-plane polarisation rotates along the radial direction, pointing toward or away from the skyrmion centre. For a Bloch-type skyrmion, $\gamma=-\frac{\pi}{2}$ and the in-plane polarisation rotates azimuthally and remains tangent to concentric circles centred on the skyrmion core. Higher-order skyrmions are obtained by increasing the integer winding number $m$, resulting in multiple rotations of the polarisation field around the skyrmion centre while preserving the radial polarisation profile.
 
Anti-Néel- and anti-Bloch-type skyrmions are obtained from the same expressions by reversing the sign of the topological order, $m \rightarrow -m$, consistent with the general classification of the anti-skyrmion as the opposite-sign topological state \cite{shen2022generation, shen2024optical, yang2025optical, shen2021topological}, while retaining the corresponding helicity ($\gamma=0$ for Néel, $\gamma=-\pi/2$ for Bloch). This reverses the winding direction of the in-plane polarisation while preserving its Néel or Bloch character, and correspondingly reverses the sign of the resulting skyrmion number.
 
Bimeron states cannot be obtained from a purely radial profile $f(r)$ alone, since a bimeron does not map the beam centre and its surroundings onto opposite poles of the Poincar\'e sphere; instead, it comprises two meron cores of opposite polarity, each covering only a hemisphere of the Poincar\'e sphere. A skyrmion texture can be converted into a bimeron by a fixed $90^\circ$ rotation of its Stokes/Poincar\'e vector about any axis lying in the $s_1$--$s_2$ plane, a freedom of the underlying topological mapping that leaves the area covered on the Poincar\'e sphere, and hence the skyrmion number, unchanged \cite{gobel2021beyond, gobel2019magnetic, shen2021topological, cisowski2023building}; the specific choice of rotation axis fixes only the orientation of the resulting bimeron within the transverse plane, not its topological class. Bimeron target fields were therefore constructed by taking the normalised Stokes vector field $\mathbf{s}(r,\phi)$ of the corresponding parent skyrmion (Néel or Bloch, $m=1$) and applying this rotation about the $s_1$-axis,
 
\begin{equation}
(s_1,s_2,s_3) \;\rightarrow\; (s_1,\,s_3,\,-s_2),
\label{eq:bimeron_rotation}
\end{equation}
 
\noindent which redistributes topological charge from the out-of-plane ($s_3$) component of the parent texture into its in-plane ($s_2$) component. The rotated Stokes field was converted back to an equivalent Jones vector via
 
\begin{equation}
\tilde{f}(r,\phi) = \arccos\!\big(s_3'(r,\phi)\big), \qquad
\tilde{\theta}(r,\phi) = \mathrm{atan2}\!\big({-}s_2'(r,\phi),\,s_1'(r,\phi)\big),
\label{eq:bimeron_angles}
\end{equation}
 
\begin{equation}
\mathbf{E}(r,\phi)
= \begin{pmatrix}
\mathbf{E}_{R}(r,\phi)\\[1ex]
\mathbf{E}_{L}(r,\phi)
\end{pmatrix}
=\begin{pmatrix} \cos\!\big(\tilde{f}(r,\phi)/2\big) \\[1ex] \sin\!\big(\tilde{f}(r,\phi)/2\big)\,e^{i\tilde{\theta}(r,\phi)} \end{pmatrix},
\label{eq:bimeron_hopf}
\end{equation}
 
\noindent where $s_1'$, $s_2'$, and $s_3'$ denote the rotated Stokes components. Unlike the Néel and Bloch type skyrmions, $\tilde{f}$ and $\tilde{\theta}$ are not separable into an independent radial profile and a linear azimuthal phase, reflecting the reduced rotational symmetry of the bimeron texture relative to its parent skyrmion.
 
The complex electric field is represented by the Jones vector
 
\begin{equation}
\mathbf{E}(x,y)
=
\begin{pmatrix}
\mathbf{E}_H(x,y)\\
\mathbf{E}_V(x,y)
\end{pmatrix}
=
\begin{pmatrix}
\frac{\mathbf{E}_R(r,\phi)+\mathbf{E}_L(r,\phi)}{\sqrt{2}}\\
-i\frac{\mathbf{E}_R(r,\phi)-\mathbf{E}_L(r,\phi)}{\sqrt{2}}
\end{pmatrix},
\end{equation}
 
\noindent where $\mathbf{E}_H(x,y)$ and $\mathbf{E}_V(x,y)$ denote the complex electric field distributions of two orthogonal polarisation components on a linear basis \cite{born1999principles, goldstein2011polarized}. Both components are multiplied by a Gaussian amplitude envelope $\exp(-r^2/2\sigma^2)$ with $\sigma = R/2$, where $R$ is the same radius as in Eq.~\eqref{eq:f_profile}, so that the envelope scales with the skyrmion and the full texture lies within the illuminated region. Because the envelope multiplies both components equally, it sets the intensity profile of the target without altering its polarisation texture or skyrmion number. These target Jones vectors are used as targets to the weighted Yang--Gu optimisation framework \cite{yang1994gerchberg, mouthaan2025generating} to determine the phase distributions displayed on the dual-SLM system, satisfying Eq. \eqref{eq:TM}.

\subsection*{Stokes parameter reconstruction}

The generated vector fields are characterised through polarisation-resolved measurements of the Stokes parameters \cite{stokes1852composition, born1999principles, goldstein2011polarized} $S_0$, $S_1$, $S_2$, and $S_3$, obtained directly from the complex-valued Jones vector components $\mathbf{E}_H(x,y)$ and $\mathbf{E}_V(x,y)$ recovered from the interferometric measurement. We define

\begin{equation}
\begin{split}
S_0(x,y) &= |\mathbf{E}_H|^2 + |\mathbf{E}_V|^2 \\
S_1(x,y) &= |\mathbf{E}_H|^2 - |\mathbf{E}_V|^2 \\
S_2(x,y) &= 2\,\mathrm{Re}\!\left(\mathbf{E}_H\mathbf{E}_V^{*}\right) \\
S_3(x,y) &= 2\,\mathrm{Im}\!\left(\mathbf{E}_H\mathbf{E}_V^{*}\right),
\end{split}
\end{equation}

\noindent and with this convention, the normalised Stokes vector is given by 

\begin{equation}
    \mathbf{s}(x,y)=\frac{(S_1(x,y),S_2(x,y),S_3(x,y))}{S_0(x,y)+\varepsilon_0}
\end{equation}

\noindent where $\varepsilon_0$ is a regularisation parameter.

\subsection*{Topological characterisation}

The topology of the generated polarisation field is quantified through the skyrmion number \cite{nagaosa2013topological, gao2020paraxial}

\begin{equation}
N_{\mathrm{sk}}
=
\frac{1}{4\pi}
\int
\mathbf{s}
\cdot
\left(
\frac{\partial \mathbf{s}}{\partial x}
\times
\frac{\partial \mathbf{s}}{\partial y}
\right)
dx\,dy.
\end{equation}

This quantity measures the number of times the polarisation field wraps around the Poincar\'e sphere \cite{shen2024optical, gao2020paraxial} and provides a quantitative verification of skyrmion formation. In practice, the integral is evaluated over the projected region after applying a Gaussian filter to reduce the noise \cite{peters2026extracting}, using the experimentally reconstructed normalised Stokes vector.

Additionally, the overlap integral (OI) between two complex fields $E_1$ and $E_2$
 
\begin{equation}
\mathrm{OI}(\mathbf{E}_1,\mathbf{E}_2)=\sqrt{\frac{\left| \iint \mathbf{E}_1(x,y) \cdot \mathbf{E}_2^*(x,y) \, dx\, dy \right|^2}{\iint |\mathbf{E}_1(x,y)|^2 \, dx\, dy \; \iint |\mathbf{E}_2(x,y)|^2 \, dx\, dy}}
\end{equation}
where $\mathbf{E}_1$ and $\mathbf{E}_2$ are the two-component Jones vector fields $(E_H, E_V)$ of Eq.~(8), so that the overlap includes both polarisation components simultaneously. OI is calculated to show the agreement between target, predicted and measured fields \cite{mouthaan2025generating}.

\backmatter

\bmhead{Data availability}
Data underlying the results presented in this paper are not publicly available at this time but may be obtained from the corresponding author upon reasonable request.

\bmhead{Code availability}
The code underlying the results presented in this paper is not publicly available at this time but may be obtained from the corresponding author upon reasonable request.

\bmhead{Supplementary information}
Supplementary Information accompanies this paper, containing Supplementary Table~1 and Supplementary Figs.~1--5.

\bmhead{Acknowledgements}
K.D. acknowledges support from the Australian Research Council (FL210100099). This research was supported by the Australian Research Council Centre of Excellence in Optical Microcombs for Breakthrough Science (project number CE230100006) and funded by the Australian Government. R.E.B.G. acknowledges support from the Australian Government Research Training Program (RTP) scholarship and from Adelaide University and Lastek Pty Ltd through the Industry Doctoral Training Centre scholarship. A.M. acknowledges support from the Australian Government Research Training Program (RTP) scholarship. All of the authors would like to acknowledge Alexander Trowbridge for creating the 3D concept diagram in Fig.~\ref{fig:concept}.

\bmhead{Author contributions}
Conceptualisation, R.E.B.G., R.M. and K.D.; experimental design and build, and experimental data acquisition, R.E.B.G.; coding, R.E.B.G., A.M. and R.M.; instrument control code, A.M.; writing---original draft, R.E.B.G.; writing---review and editing, R.E.B.G., R.M., A.M. and K.D.; supervision, R.M. and K.D.; resources, K.D.

\bmhead{Competing interests}
The authors declare no competing interests.

\bibliography{references}

\clearpage
\section*{Figures}

\clearpage

\clearpage

\clearpage

\clearpage

\end{document}